\documentclass[aps,physrev,reprint,groupedaddress]{revtex4-2}

\usepackage{amsmath,amsfonts}
\usepackage{graphicx} 
\usepackage{xcolor}
\usepackage{comment}
\usepackage{placeins} 
\usepackage[hyperindex,breaklinks]{hyperref}
\hypersetup{linkcolor=blue, citecolor=red, colorlinks=true}
\usepackage{url}
\usepackage{capt-of}
 \usepackage{dcolumn} 
\usepackage{bm} 
\usepackage{soul}

\begin{document}
  
\title{Composing $\alpha$-Gauss and logistic maps: Gradual and sudden transitions to chaos}
  
\author{Marcelo A. Pires$^{1}$}
\email{piresma@cbpf.br}

\author{Constantino Tsallis$^{1,2,3,4}$}
\email{tsallis@cbpf.br}

\author{Evaldo M.F. Curado$^{1,2}$}
\email{evaldo@cbpf.br}

\affiliation{%
$^{1}$ Centro Brasileiro de Pesquisas Fisicas, 
Rua Dr. Xavier Sigaud 150, 22290-180, Rio de Janeiro, Brazil 
\\
$^{2}$ National Institute of Science and Technology for Complex Systems, Rua Dr. Xavier Sigaud 150, 22290-180, Rio de Janeiro, Brazil 
\\
$^{3}$ Santa Fe Institute, 1399 Hyde Park Road, 87501, Santa Fe, USA
\\
$^{4}$ Complexity Science Hub Vienna, 
 Metternichgasse 8, 1030 Vienna, Austria
}%

\date{\today}

\begin{abstract}
We introduce the $\alpha$-Gauss-Logistic map, a new nonlinear dynamics constructed by composing the logistic and $\alpha$-Gauss maps.  Explicitly, our model is given by  
$x_{t+1} =  f_L(x_t)x_t^{-\alpha}  -   \lfloor  f_L(x_t)x_t^{-\alpha} \rfloor $  
where $f_L(x_t) = r x_t (1-x_t)$ is the logistic map and $ \lfloor \ldots \rfloor  $ is the integer part function.
Our investigation reveals a rich phenomenology depending solely on two parameters, $r$ and $\alpha$. 
For $\alpha < 1$, the system exhibits 
\textcolor{black}{multiple} period-doubling cascades to chaos as the parameter $r$ is increased, interspersed with stability windows within the chaotic attractor. In contrast, for $1 \leq \alpha < 2$, the onset of chaos is abrupt, occurring without any prior bifurcations, and the resulting chaotic attractors emerge without stability windows. 
For $\alpha \geq 2$, the regular behavior is absent.
The special case of $\alpha = 1$ allows an analytical treatment, yielding a closed-form formula for the Lyapunov exponent and conditions for an exact uniform invariant density, using the Perron-Frobenius equation. Chaotic regimes for $\alpha = 1$ can exhibit gaps or be gapless.  
Surprisingly, the golden ratio $\Phi$ marks the threshold for the disappearance of the largest gap in the regime diagram. 
Additionally, at the edge of chaos in the abrupt transition regime, the invariant density approaches a $q$-Gaussian with $q=2$, which corresponds to a Cauchy distribution.  
\end{abstract}

 \maketitle


\section{Introduction} \label{sec:intro}

\begin{figure*}[!htb]
    \centering
    \includegraphics[scale=0.35]{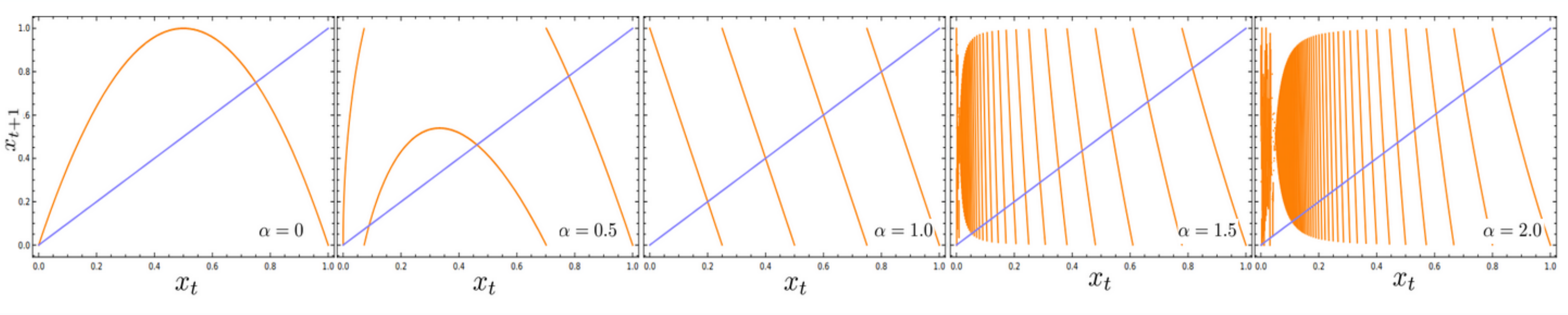}
    \caption{Return map of the $\alpha$-Gauss-logistic  ($\alpha$GL) map for $r=4$.}
    \label{fig:return_map}
\end{figure*}

The study of nonlinear dynamical systems has revealed fundamental mechanisms through which simple deterministic equations can produce complex, chaotic behavior~\cite{strogatz2024nonlinear,ausloos2006logistic}. Notably, chaos theory holds significant value not only for its theoretical insights but also for its  applications~\cite{grassi2021chaos,hamidouche2024mastering,mashuri2024application}.

The logistic map~\cite{may1976simple} serves as one of the most paradigmatic  examples of dissipative nonlinear dynamics. It is governed by the quadratic recurrence relation
\begin{align}
x_{t+1} &= r x_t (1 - x_t), \label{eq:logistic_map}
\end{align}
where $x_t \in [0, 1]$ represents the state at time $t=0,1,2,\dots$, and $r \in [0, 4]$ is the control parameter. This deceptively simple map exhibits remarkable dynamics, including the now-classic period-doubling route to chaos, where increasing $r$ leads to successive bifurcations from stable fixed points to periodic orbits and eventually to chaotic behavior.

Recent work~\cite{Beck2024} has introduced the $\alpha$-Gauss map defined by
\begin{align}
x_{t+1}=\frac{1}{x_t^\alpha} - \Bigl\lfloor\frac{1}{x_t^\alpha} \Bigr\rfloor, \label{eq:alpha_gauss_function}
\end{align}
where $x_t \in [0, 1]$, $\alpha \ge 0$, and $\lfloor\cdot\rfloor$ denotes the floor function. This map reduces to the standard Gauss map when $\alpha = 1$~\cite{adler1984backward, sinai2008renewal, corless1992continued, banchoff1982cusps, BeckSchlogel1994thermodynamics} (also known as continued fraction map).  The $\alpha$-Gauss map displays qualitatively different behavior from the logistic map, with an abrupt transition to chaos~\cite{Beck2024} that occurs without the intervening period-doubling cascade characteristic of the logistic map.

While the logistic map clarifies how chaos can emerge gradually through an infinite sequence of period-doubling cascades, the $\alpha$-Gauss map shows that chaos can also appear suddenly, without intermediate bifurcations, as a parameter crosses a critical point. This dichotomy gradual-nongradual motivates our central research question: Can these disparate routes to chaos be unified within a single theoretical framework?

While previous works~\cite{
kawabe1991fractal,alvarez2008critical,kawabe1991intermittent,pikovsky1983intermittent,cosenza2010lyapunov,schuster2006deterministic,aguirregabiria2009robust,Banerjee1998,ZeraouliaSprott2012,AndrecutPRE2001,gallas2010structure} have examined non-gradual transitions to chaos in various contexts, these studies focused on different classes of maps. 
Furthermore, the absence of stability islands in chaotic attractors - a characteristic feature of {\em robust chaos} \cite{Banerjee1998,ZeraouliaSprott2012} - has been established for discrete-time formulations~\cite{aguirregabiria2009robust,Banerjee1998,ZeraouliaSprott2012,AndrecutPRE2001} as well as for continuous-time systems~\cite{gallas2010structure}.

Before proceeding, note that our proposal is to be distinguished from the composed
Logistic-Gauss map~\cite{da2021logistic,de2025dynamical}. 
While their mathematical approach presents an interesting alternative formulation, it employs the Gaussian map $x_{t+1} = e^{-\alpha x_t^2} + \beta$ in conjunction with a logistic-like model, rather than utilizing a generalization of the Gauss continued fraction map~\cite{Beck2024} as in our present work.

The manuscript is organized as follows: Section~\ref{sec:model} describes our unified model; 
\textcolor{black}{Section~\ref{sec:general} presents 
general results; 
Section~\ref{sec:alpha0} focuses on the case with $\alpha=0$};   
Section~\ref{sec:alpha1} examines the special case $\alpha = 1$; Section~\ref{sec:jump} analyzes  the jump to chaos and the edge of chaos; and Section~\ref{sec:final} offers concluding remarks.

\section{The  $\alpha$-Gauss-logistic  Map}\label{sec:model}

To address the main question we posed in the previous section, we introduce a new model by composing the logistic map with the $\alpha$-Gauss map. The model is defined as follows:
\begin{align}
f_L(x_t) &= r x_t (1-x_t)
\\
x_{t+1}  &=  f_L(x_t)x_t^{-\alpha} -  \lfloor  f_L(x_t)x_t^{-\alpha} \rfloor
 \,.
\end{align}
Equivalently, 
\begin{align}
g(x_t) &= r x_t^{1-\alpha} (1-x_t), \label{eq:g_function} \\
x_{t+1} &= f(x_t) = g(x_t) - 
\lfloor g(x_t \rfloor, \label{eq:alpha_gauss_logistic_map}
\end{align}
where $x_t \in [0, 1]$, $\alpha \ge 0$, and $r$ is a parameter that is no longer restricted to the interval $[0, 4]$.
We name this dynamical system the $\alpha$-Gauss-Logistic 
(or just $\alpha$GL).

\begin{figure*}[!htb]
\centering

\includegraphics[scale=0.673]{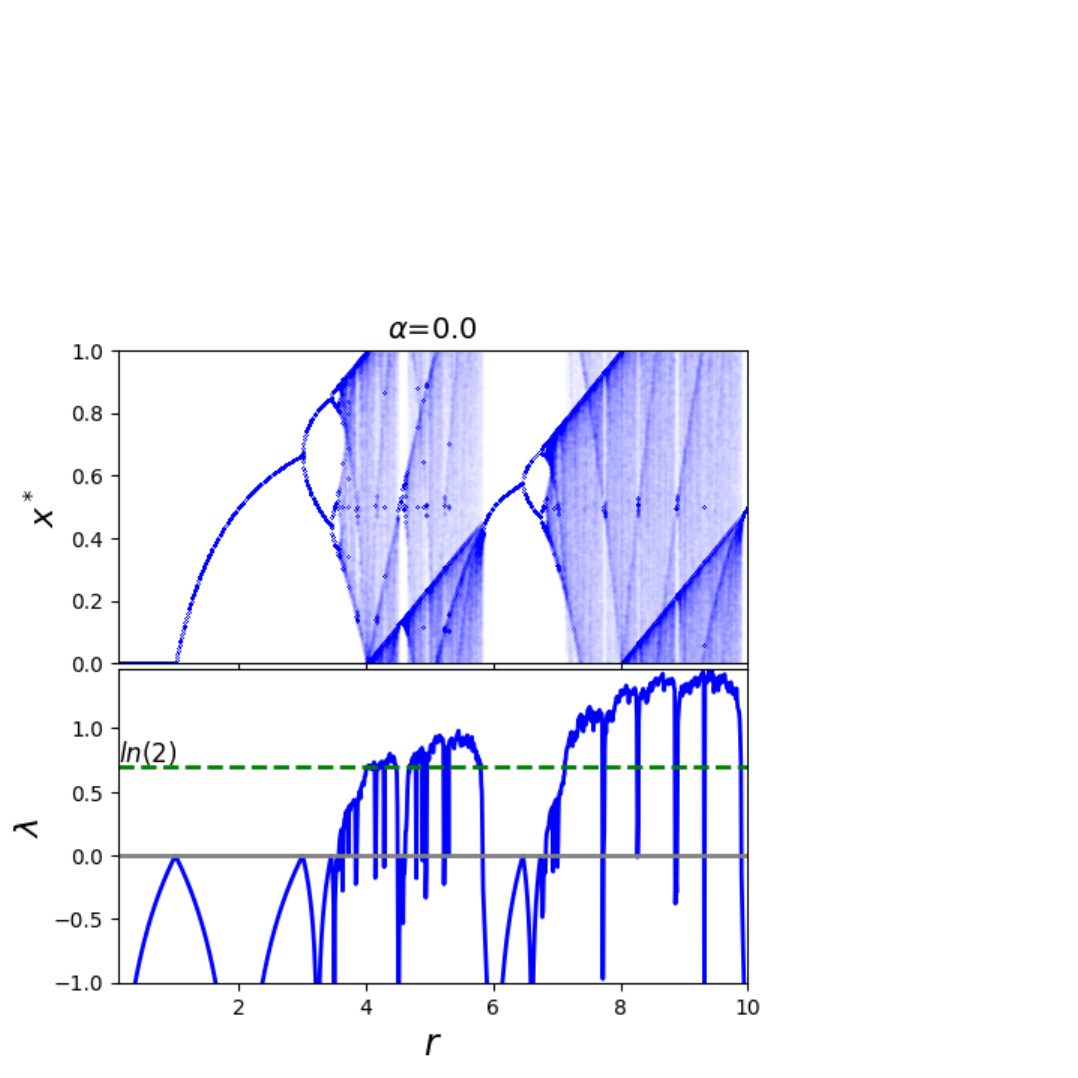}
\includegraphics[scale=0.673]{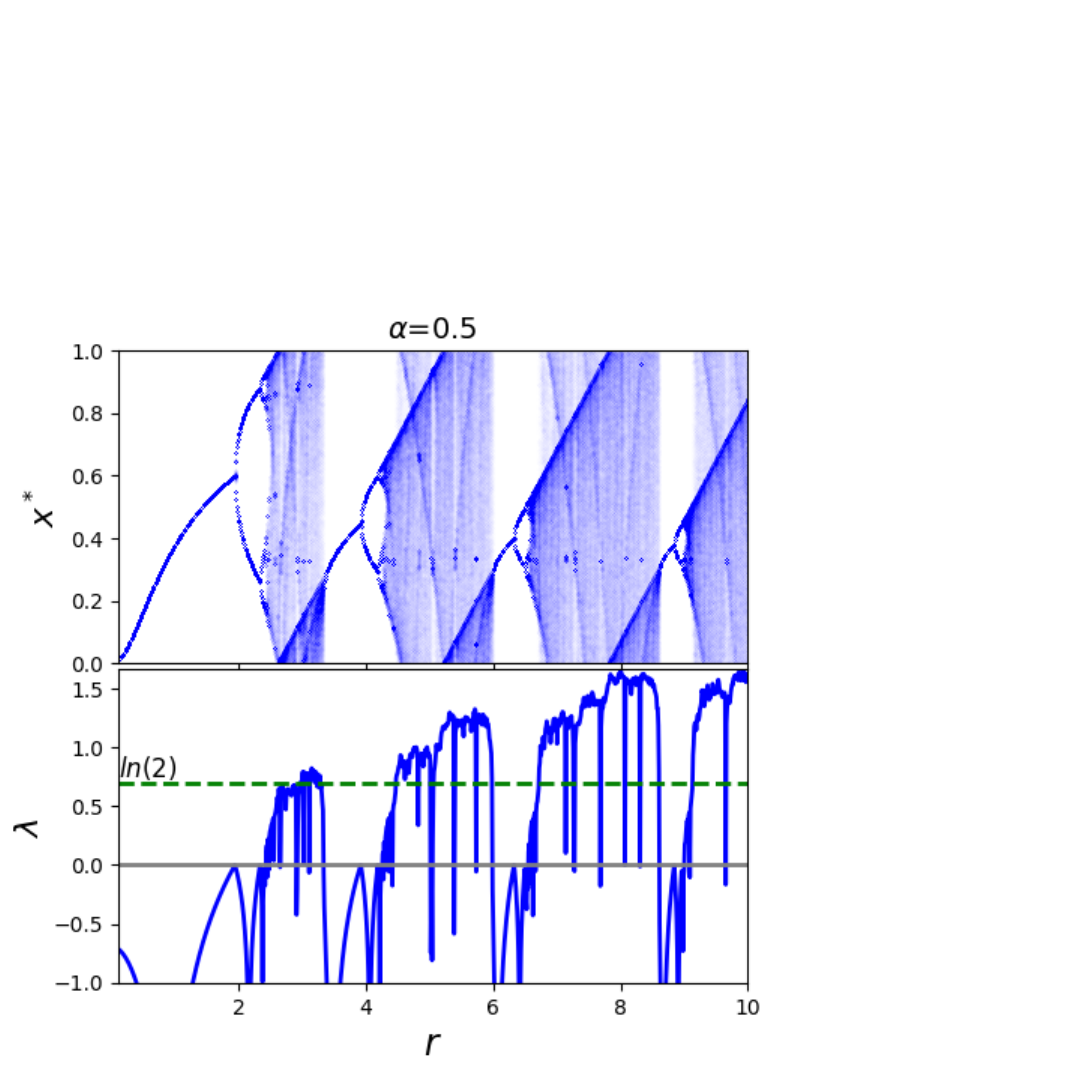}

\includegraphics[scale=0.673]{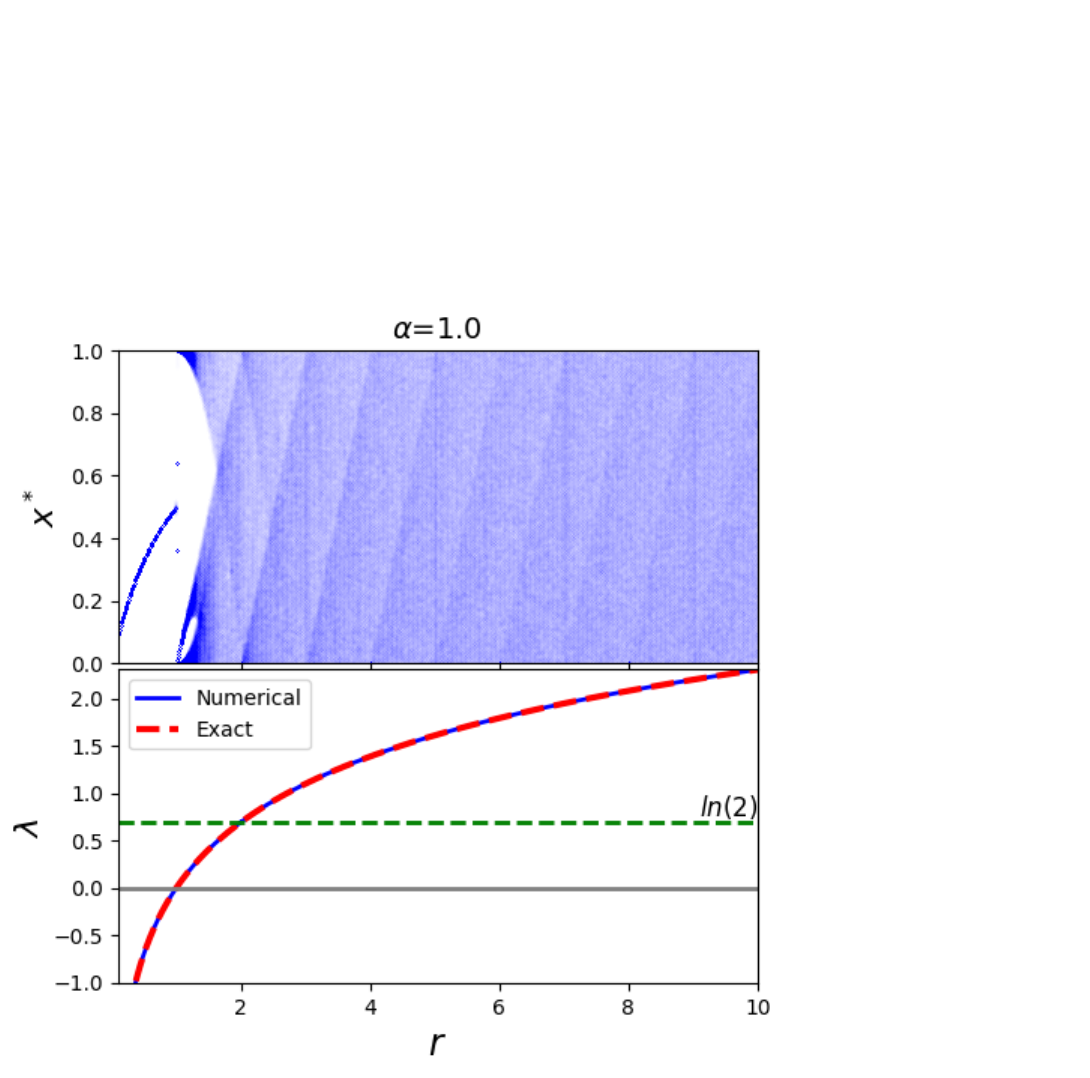}
\includegraphics[scale=0.673]{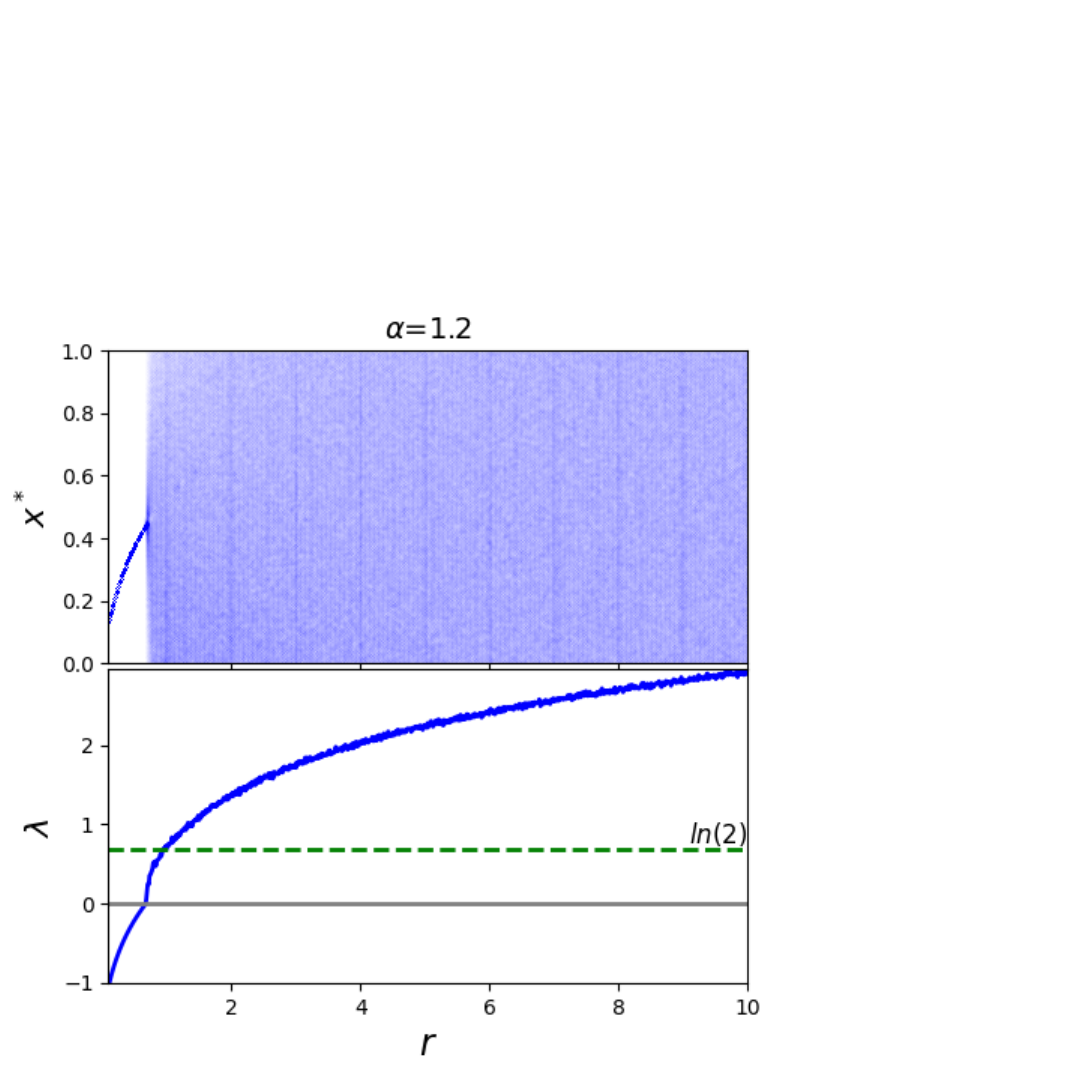}
    \caption{Bifurcation diagrams (top panels) and corresponding Lyapunov exponents (bottom panels) for the $\alpha$GL map as a function of the parameter $r$, for key values of $\alpha$. For comparison, the horizontal green line indicates $ \ln(2)$, the maximum $\lambda$ that is obtained for the standard logistic map. \textcolor{black}{Note that for $\alpha<1$ there are multiple bifurcation cascades and stable windows. } }
    \label{eq:bif_lyap}
\end{figure*}

\begin{figure}[!htb]
    \centering
    \includegraphics[scale=0.49]{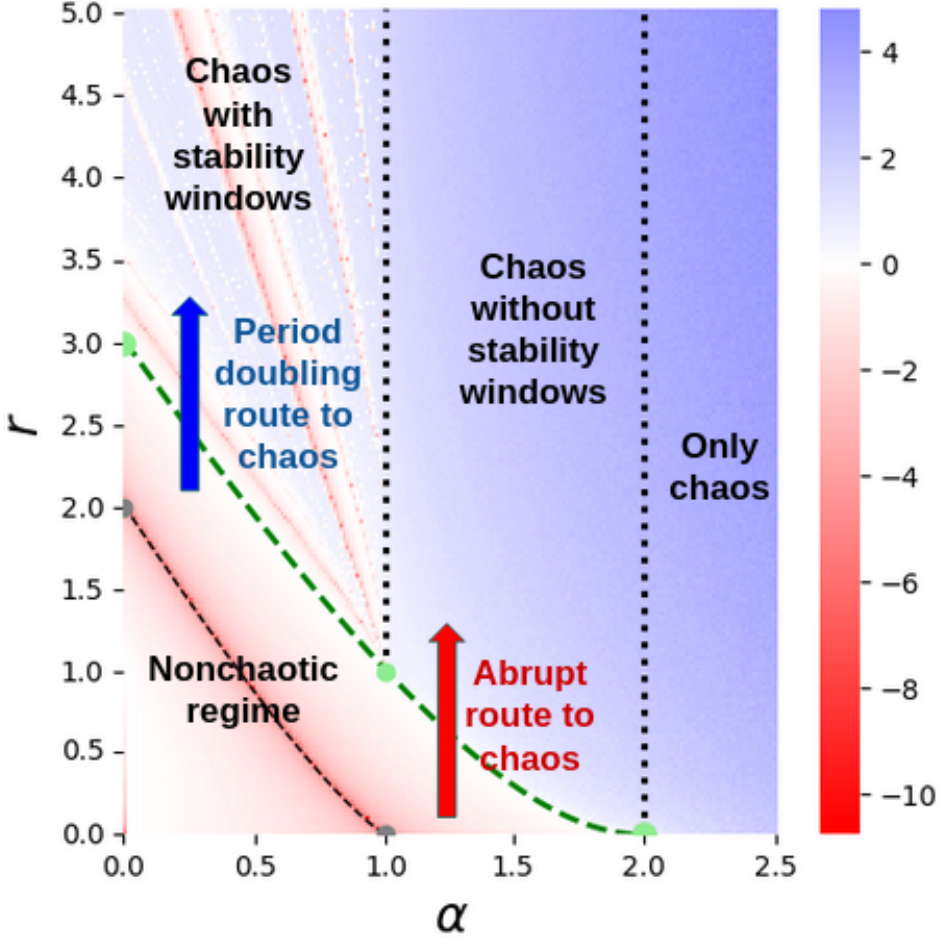}
    \caption{Characterization of the dynamical  regimes present in the $\alpha$GL map. The colors represent the value of the Lyapunov exponent, $\lambda$, obtained numerically from  \eqref{eq:lyapunov_exponent_expanded}. The arrows indicate the routes through which the system transitions to chaos. \textcolor{black}{The  critical curve (green) that passes through the points $\{(0,3),(1,1),(2,0)\}$ comes
     from the analytical solution \eqref{eq:crit_r_exact}.
     In turn, the superstable curve (black) passing through the points $\{(0,2),(1,0)\}$ comes from the exact equation~\eqref{eq:superstable_r}.
     }  }
    \label{fig:phase_diagram}
\end{figure}

\begin{figure*}[!htb]
\centering
\includegraphics[scale=0.49]{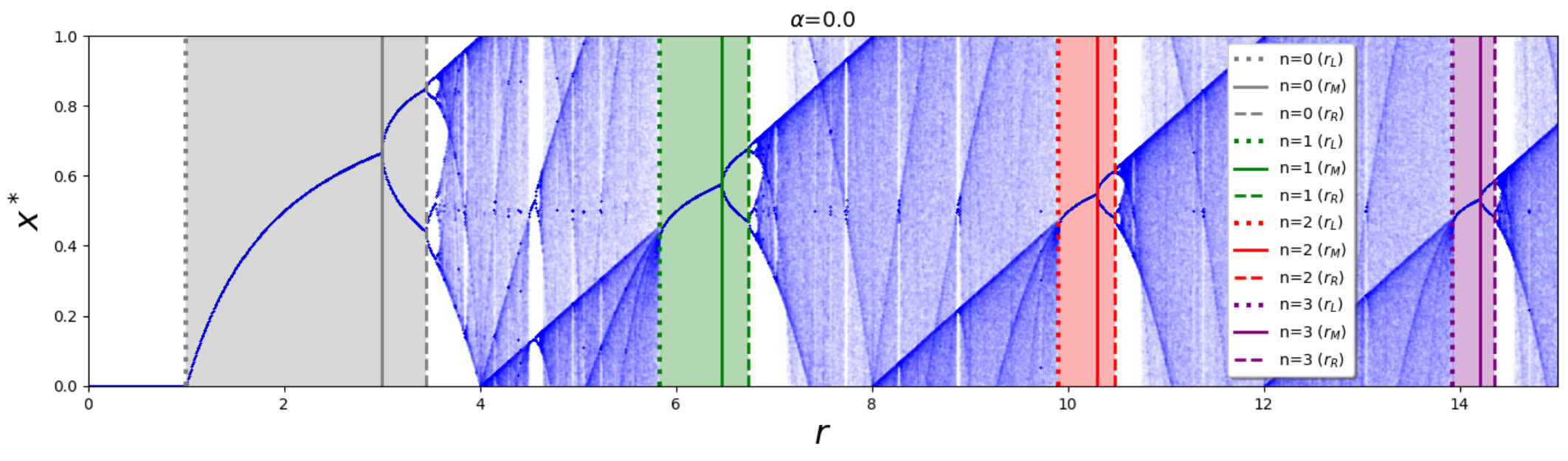}
\caption{
\textcolor{black}{Multiple sequential bifurcation cascades in the extended logistic map  ($\alpha$GL with $\alpha=0$).
The vertical lines delimit exactly the range of $r$ values where there are fixed points and cycles-2 as established by 
Eqs.~(\ref{eq:fixedp_onset}-\ref{eq:cycle2_onset}).
}
}
\label{fig:bif_alp0}
    \includegraphics[scale=0.42]{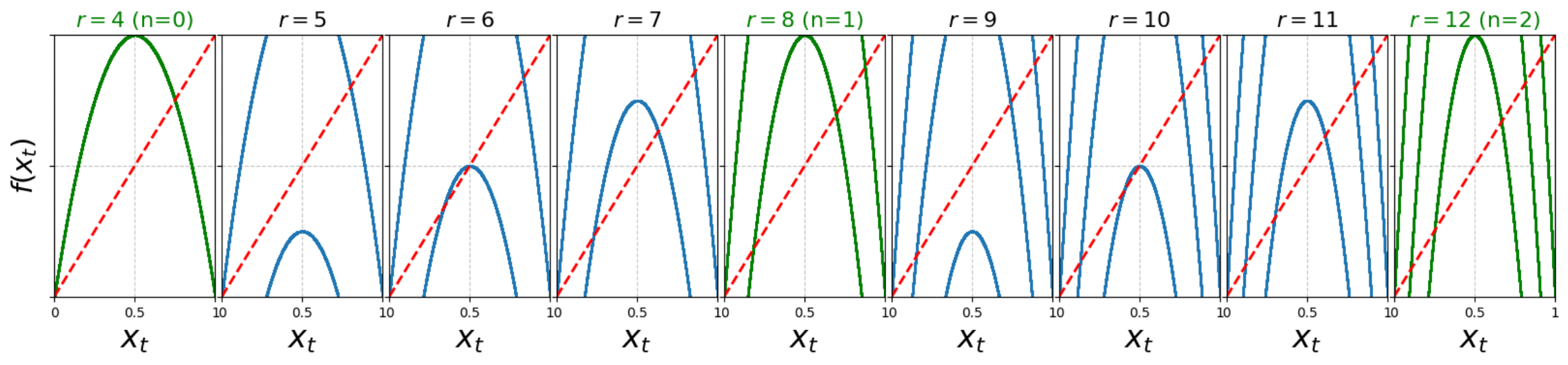}
\caption{
\textcolor{black}{Return map of the extended logistic map. Every new parabola emerges rightly after $r_p = 4(n+1) $ 
 as established in \eqref{eq:parabola_onset}. 
}
}
\label{fig:return_map_alp0}
\end{figure*}

Figure~\ref{fig:return_map} illustrates representative cases for the  return map of the $\alpha$GL model. When $\alpha = 0$, the return map  shows the characteristic parabolic curve of the logistic map. At $\alpha=1$, it reduces to a piecewise linear function, producing a sawtooth-like pattern. Other $\alpha$ values result in maps with various non-trivial features.

\section{General case}\label{sec:general}
 
\subsection{Lyapunov Exponent}

The Lyapunov exponent $\lambda$ measures the rate of exponential divergence of nearby trajectories and it is given by:
\begin{align} \label{eq:lyapunov_definition}
    \lambda = \lim_{T \to \infty} \frac{1}{T} \sum_{t=0}^{T-1} \ln \left| f'(x_t) \right|,
\end{align}
where $f'(x_t)$ is the derivative of the map $f(x_t)$ with respect to $x_t$.

If $\lambda > 0$, nearby trajectories tend to diverge exponentially, indicating chaotic behavior. If $\lambda < 0$, nearby trajectories tend to converge exponentially, indicating stability. If $\lambda = 0$, the system is marginally stable. 

Figure~\ref{eq:bif_lyap} displays  bifurcation diagrams (top panels) and Lyapunov exponents (bottom panels) for 
representative scenarios in the $\alpha$GL map. Remarkably, when $\alpha = 0$  and $r>4$, the $\alpha$GL map already exhibits greater chaoticity than the standard logistic map, since there are regions with $\lambda> \ln 2 $ (above the green line). 
With rising $\alpha$, the period-doubling cascade weakens, disappearing at $\alpha=1$. Beyond this threshold, the system undergoes an abrupt transition to chaos, and the Lyapunov exponent shifts from nonmonotonic to monotonic behavior. 

\color{black}
In order to better understand the results in Fig.~\ref{eq:bif_lyap}, first note that the derivative of $f(x_t)$ can be expressed as $ f'(x_t) = g'(x_t)$
since the floor function  $\lfloor g(x_t) \rfloor$  is piecewise constant. Then, \eqref{eq:lyapunov_definition} becomes
\begin{align} \label{eq:lyapunov_exponent_expanded}
\lambda &= \ln{r} + \psi(\alpha) \\
\psi(\alpha) &= \lim_{T \to \infty} \frac{1}{T} \sum_{t=0}^{T-1} \ln \left|  
\frac{(1 - \alpha)}{x_t^{\alpha}}
- 
\frac{(2 - \alpha)}{x_t^{\alpha-1}}
\right|.
\end{align}

This decomposition $\lambda = \ln{r} + \psi(\alpha)$ indicates that the
transition from nonmonotonic behavior  ($\alpha<1$) to 
monotonic ($\alpha\geq1$) comes from the competition between: (i) a monotonic contribution related to $\ln{r}$; (ii) a nonmonotonic  contribution related to $\psi(\alpha)$. At the transition point $\alpha=1$   there is no nonmonotonic contribution since $\psi(1)=0$.
\color{black}

Figure~\ref{fig:phase_diagram} clarifies the regime diagram of the $\alpha$GL map. 
Three main scenarios are observed: 
\begin{itemize}
\item (i) multiple period-doubling cascades for $\alpha < 1$,
\item  (ii) an abrupt transition from a nonchaotic regime to chaos for $1 \leq \alpha < 2$, and 
\item (iii) absence of regular dynamics for $\alpha \geq 2$.
\end{itemize}

\subsection{Fixed points}

The fixed points $x^*$ satisfy:
\begin{align}
    x^* &= g(x^*) - n, \label{eq:fixed_point_condition}
\end{align}
where $n^* =  \lfloor g(x^*) \rfloor  \in \mathbb{Z}$.  
Substituting the expression for $g(x)$ gives:
\begin{align}
    r (x^*)^{1-\alpha} (1 - x^*) &= x^* + n^*, \label{eq:explicit_fixed_point}
\end{align}

Finding analytical solutions for the above fixed-point equation  generically depends on the values of $\alpha$ and $r$.

\color{black}
First, let us obtain an analytical solution for the case with  $n^*=0$. 
In this case, \eqref{eq:explicit_fixed_point} provides 
\begin{align} \label{eq:crit_r}  
    r^* = \frac{(x^*_c)^{\alpha}}{1 - x^*_c}.
\end{align}
Substituting \eqref{eq:crit_r} into the stability condition for a fixed point, $f'(x^*_c) = -1$, we obtain
\begin{align}
-1 &= f'(x^*_c) = 
\frac{(x^*_c)^{\alpha}}{1 - x^*_c} 
\left[ 
\frac{1-\alpha}{(x^*_c)^{\alpha} }
- 
\frac{2-\alpha}{(x^*_c)^{\alpha-1}}
 \right] \label{eq:fixp_intermedi}
\end{align}
Solving \eqref{eq:fixp_intermedi}, we obtain
\begin{align} \label{eq:crit_x}   
    x^{*}_c = \frac{2 - \alpha}{3 - \alpha}.
\end{align}
Substituting \eqref{eq:crit_x} into \eqref{eq:crit_r}, we arrive at
\begin{align}  \label{eq:crit_r_exact}  
 r^*(\alpha) = (2 - \alpha)^{\alpha} (3 - \alpha)^{1-\alpha}.    
\end{align}
Figure~\ref{fig:phase_diagram} shows that
this closed-form formula accurately captures 
the critical line passing through the points
$r^*(0) = 3$, $r^*(1) = 1$ and $  r^*(2) = 0$. 

\color{black}

\subsection{Superstability}
\color{black}
A fixed point $x^*_s$ of a 1D map exhibits superstability when two conditions are simultaneously met: its derivative at that point is zero, i.e., $f'(x^*_s) = 0$, and it is a fixed point, $x^*_s = f(x^*_s)$. For  $\alpha$GL these conditions yield the following closed-form solutions:
\begin{align}    
    x^*_s &= \frac{1-\alpha}{2-\alpha} \\
    r_s &= (1-\alpha)^\alpha(2-\alpha)^{1-\alpha}.
\label{eq:superstable_r}
\end{align}
While the Lyapunov exponent, as defined in \eqref{eq:lyapunov_definition}, theoretically approaches $-\infty$ at superstable points, numerical computations yield finite values. 
As illustrated in Fig.~\ref{fig:phase_diagram}, the analytical expression for $r_s$ presented in \eqref{eq:superstable_r} effectively delineates the curve of minimal (most negative) Lyapunov exponents, traversing the points where $r=2$ at $\alpha=0$ and $r=0$ at $\alpha=1$

\color{black}

\color{black} 
\section{Special Case I: The extended logistic map}\label{sec:alpha0}

When  $\alpha=0$,  we obtain
\begin{align}
x_{t+1} =  r x_t (1-x_t) - 
\lfloor r x_t (1-x_t) \rfloor.
\label{eq:extended_logistic}
\end{align}

We call this new model the \emph{extended logistic map} because it is a logistic-like map with a parameter that does not require anymore $r \in [0,4]$.  
This model has not been explored in previous generalizations of the logistic map~\cite{radwan2013some,borujeni2015modified,lawnik2017generalized,da2017route,sayed2017generalized,leonel2019allee,hamada2025investigating,zhang2025chaos,abdellah2025generalized}.

In short notation we have $x_{t+1} =  f_L(x_t) - n$ where $n = \lfloor f_L(x_t) \rfloor $.  
At the fixed-point,  \eqref{eq:extended_logistic} becomes
\begin{equation}
    r x^{*2} + (1 - r)x^* + n = 0. \label{eq:quadratic_form}
\end{equation}

From   \eqref{eq:quadratic_form} the discriminant condition $\Delta \geq 0$ yields  the values of $r$ for which the fixed points appear for each $n$,
\begin{align}
r_L=1 + 2 n + 2 \sqrt{n^2+n}
\label{eq:fixedp_onset}
\end{align}

The values of $r$ where the fixed points lose stability, 
$ f'(x^{*})=-1 $, for each $n$ are
\begin{align}
r_M = 1 + 2 n + 2 \sqrt{n^2+n+1}
\label{eq:fixedp_end}
\end{align}

Now let us focus on cycle-2. The values of $r$ where the cycle-2 appears for each $n$ is $r_M$ which is the same value where the fixed-point loses stability. 
On the other hand, the values of $r$ where the cycle-2  loses stability for each $n$ occurs 
when $f'(x^{*}_1) f'(x^{*}_2) = -1$, which implies that
\begin{align}
r_R= 1 + 2 n + \sqrt{2} \sqrt{2 n^2+2 n+3}
\label{eq:cycle2_onset}
\end{align}

Figure~\ref{fig:bif_alp0} shows that Eqs.~(\ref{eq:fixedp_onset}-\ref{eq:cycle2_onset}) are in good agreement with the simulations. Additionally, see that the extended logistic map presents  several bifurcation cascades. 

The size of the stable fixed-point window for each $n$ is asymptotically described by
\begin{align}
r_M-r_L \sim
\frac{1}{n} -\frac{1}{2 n^2}+
O\left(\left(\frac{1}{n}\right)^3\right)
\label{eq:fixedp_region}
\end{align}

The size of the stable cycle-2 window for each value of $n$ is asymptotically described by 
\begin{align}
r_R-r_M \sim  
\frac{1}{2 n}-\frac{1}{4 n^2} + 
O\left(\left(\frac{1}{n}\right)^3\right)
\label{eq:cycle2_region}
\end{align}

Equations~(\ref{eq:fixedp_region}-\ref{eq:cycle2_region}) show that 
as $n$ increases the region with fixed-points and cycle-2 decreases, as indeed observed in Fig.~\ref{fig:bif_alp0}.

Figure~\ref{fig:return_map_alp0} shows that a new branch appears for multiples of $4$. Why?
First, recall that for $x \in [0,1]$, the logistic map reaches its maximum at $x_m = 1/2$. 
A new branch appears when  the peak of $f_L(x)$ surpasses an integer.
Then, 
a new parabola appears when 
$r_p/4 = n+ 1$ which implies that 
\begin{align}
r_p = 4 (n+ 1).
\label{eq:parabola_onset}
\end{align}
This equation 
matches the results shown in Fig.~\ref{fig:return_map_alp0} 
where the number of parabolas depends on $n = \lfloor f_L(x) \rfloor$.

\color{black} 

\section{Special Case II: The $r$-map }\label{sec:alpha1}

For $\alpha=1$ we have the nonlinear recurrence equation 
\begin{align}
x_{t+1} = g(x_t) = r - r x_t - \lfloor r - r x_t \rfloor. \label{eq:rmap}
\end{align}
\textcolor{black}{As this particular map only depends on $r$ we name it \emph{$r$-map}.}

\subsection{Lyapunov exponent}

From \eqref{eq:rmap} we have $g'(x_t) = -r$. Then, the Lyapunov exponent of $r$-map is
\begin{align} \label{eq:lyapunov_alpha1}
    \lambda_{\alpha=1} = \ln r.
\end{align}
\textcolor{black}{This logarithmic relationship demonstrates that chaos in the $r$-map  increases monotonically with $r$, exhibiting no stability windows - a distinctive feature when compared to the logistic map.} Moreover, from \eqref{eq:lyapunov_alpha1} we obtain the exact location of the transition point at $(\alpha,r)=(1,1)$, as indicated in Fig.~\ref{fig:phase_diagram}.

\subsection{Analytical temporal solution for $r<1$}

For $r<1$, we obtain $g(x_t) = r x_t (1-x_t) < 1$ 
since $x_t \in [0, 1]$. Then $n = 0$. 
This condition reduces \eqref{eq:rmap} to an exactly solvable model expressed by a first-order linear recurrence relation that can be written as
\begin{align}
x_{t+1} + r x_t = r. \label{eq:non-homogeneous}
\end{align}

The homogeneous solution $x_t^{(h)}$ satisfies the equation $x_{t+1}^{(h)} + r x_t^{(h)} = 0$, which has a solution in the form
\begin{align}
x_t^{(h)} = A (-r)^t, \label{eq:homogeneous_solution}
\end{align}
where $A$ is an arbitrary constant.

For a  particular case $x_t^{(p)}$, \eqref{eq:non-homogeneous} implies that we can assume a constant solution of the form $x_t^{(p)} = C$. Substituting this into the nonhomogeneous \eqref{eq:non-homogeneous}:
\begin{align}
C + r C = r \Rightarrow   C = \frac{r}{1 + r}. \label{eq:particular_solution}
\end{align}
Thus, the particular solution is:
\begin{align}
x_t^{(p)} = \frac{r}{1 + r}.
\end{align}
The general solution is the sum of the homogeneous and particular solutions:
\begin{align}
x_t = x_t^{(h)} + x_t^{(p)} = A (-r)^t + \frac{r}{1 + r}. \label{eq:general_solution}
\end{align}
Using the initial condition $x_0$ at $t=0$:
\begin{align}
x_0 = A (-r)^0 + \frac{r}{1 + r} \Rightarrow
A = \frac{x_0 + rx_0 - r}{1 + r}. \label{eq:constant_A}
\end{align}
Substituting the value of $A$ from \eqref{eq:constant_A} into the general solution \eqref{eq:general_solution}, we obtain that the exact solution to  \eqref{eq:non-homogeneous}  is given by
\begin{align}
x_t = \left( \frac{x_0 + rx_0 - r}{1 + r} \right) (-r)^t + \frac{r}{1 + r}. \label{eq:exact_solution}
\end{align}

\begin{figure}[!htb]
    \centering
    \includegraphics[width=0.45\textwidth]{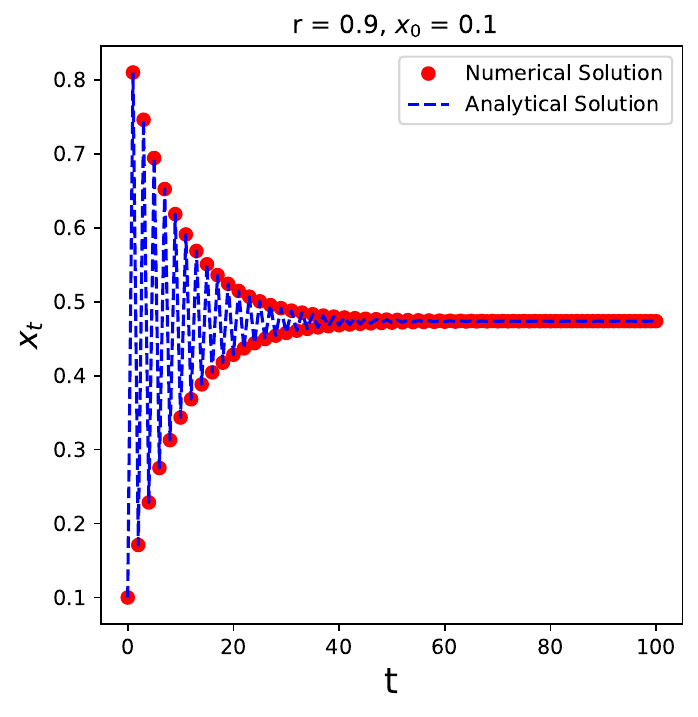}
    \caption{Exact solution, \eqref{eq:exact_solution}, for the condition $r<1$ and $\alpha=1$.}
    \label{eq:aGL_rless1_alp1}
\end{figure}

Figure~\ref{eq:aGL_rless1_alp1} shows that this closed-form solution agrees well with the simulations.

\begin{figure}[h]
\centering
\includegraphics[width=0.499\textwidth]{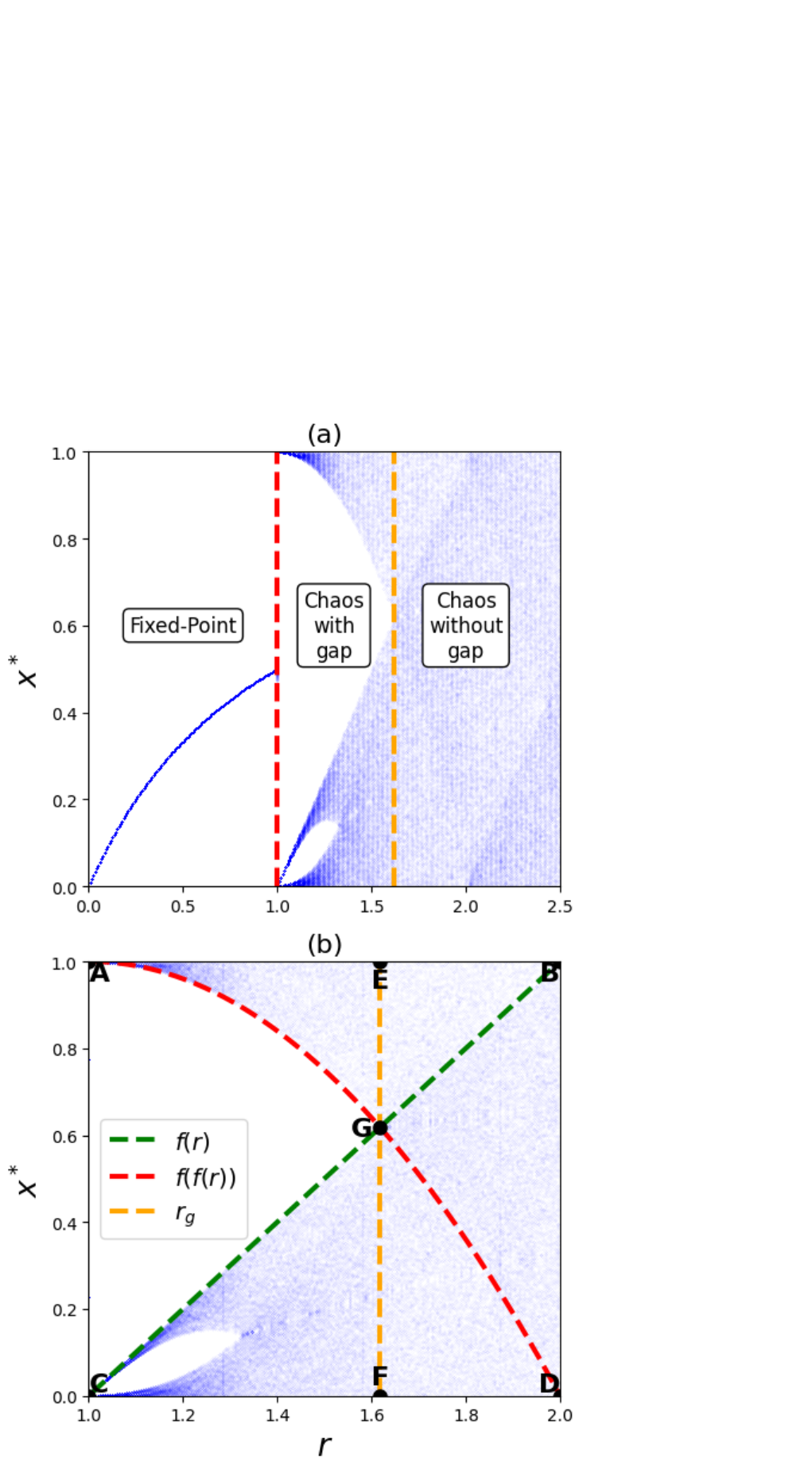}
\caption{  Long-term behavior of the $\alpha$GL map for $\alpha=1$.  
(a) Regime diagram where the gaps correspond to values of $x_t$ that the system never visits after an  initial transient. (b) Zoom at the  region with gaps where the curves for $f(r)$  and $f(f(r))$ are also plotted. The orange line indicates that our theoretical calculation of the end of the largest gap $r_g$ agrees well with the simulations.}
\label{fig:agl_largest_gap}
\end{figure}

\begin{figure*}[!htb]
\centering
\includegraphics[width=0.987\textwidth]{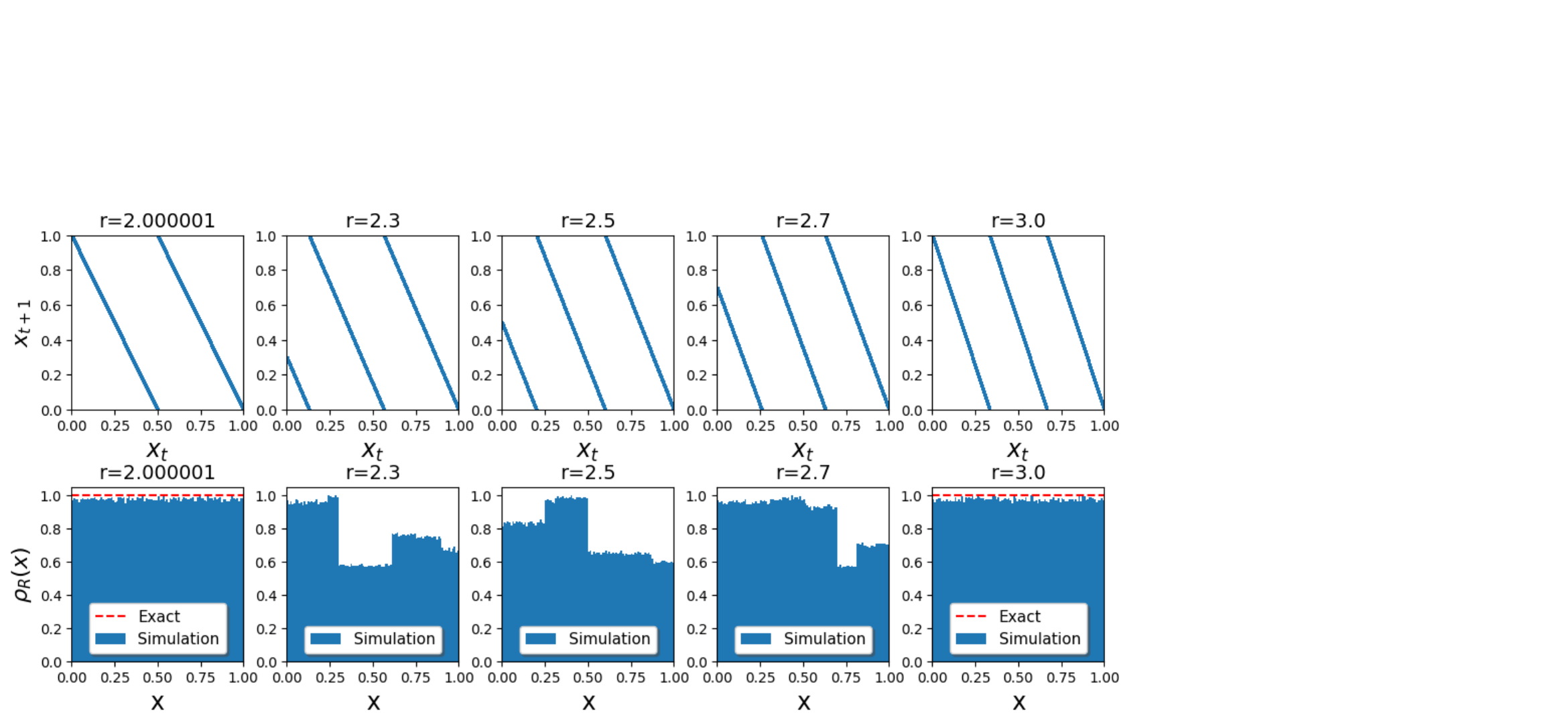}
\caption{Return map (upper panels) and corresponding relative invariant density $\rho_R(x) = \rho(x)/\rho_{max} $ (lower panels) for $\alpha=1$ and increasing values of $r$. 
\textcolor{black}{The case $r=2+\delta$ with $\delta=10^{-6}\ll1$ and $r=3$ lead to a uniform invariant density as established in \eqref{eq:rho_unif_alp1}. } } 
\label{fig:agl_return_and_density_alp1}
\end{figure*}

\subsection{Analytical stationary solution for $r<1$}

To analyze the long-term behavior of the solution given by   \eqref{eq:exact_solution} we consider the limit of $x_t$ as $t \to \infty$. 

Since we consider the case where $r < 1$ then $\lim_{t \to \infty} (-r)^t = 0$. Thus, the long-term behavior of the system is given by 
\begin{align}
x^* = \lim_{t \to \infty} x_t = \frac{r}{1 + r}. \label{eq:long_term_behavior}
\end{align}
This shows that for $r < 1$, the solution $x_t$ converges to the fixed point $x^* = \frac{r}{1 + r}$ as $t$ approaches infinity, regardless of the initial condition $x_0$.

\subsection{On the largest gap}

Figure~\ref{fig:agl_largest_gap} shows that the largest gap starts from $r=1$ and ends at a point $r_g$ determined by the intersection of the functions $f(r,x)$ and $f(r,f(r,x))$, which are given by
\begin{align} \label{eq:f}
    f(r,x) &= r(1-x) - \lfloor r(1-x) \rfloor, \\
    f(r,f(r,x)) &= r(1-f(r,x)) - \lfloor r(1-f(r,x)) \rfloor.
\end{align}
Since $f(r,x)$ varies from 0 to 1 when $r$ varies from 1 to 2, the point $x=0$ is relevant to determine the boundary of the gap. Thus, we consider $f(r) \equiv f(r, 0) = r - \lfloor r \rfloor$. For $r \in (1,2)$, we have $\lfloor r \rfloor = 1$, yielding:
\begin{align} \label{eq:interval}
    f(r) &= r - 1, \\
    f(f(r)) &= r(2 - r) - \lfloor r(2 - r) \rfloor.
\end{align}

At the intersection point where $f(r) = f(f(r))$, we have:
\begin{align} \label{eq:intersection}
    r - 1 &= r(2 - r) - \lfloor r(2 - r) \rfloor.
\end{align}

For $r \in (1,2)$, the function $r(2-r)$ decreases from 1 to 0, which implies $\lfloor r(2 - r) \rfloor = 0$. In this case,  \eqref{eq:intersection} simplifies to:
\begin{align} \label{eq:simplified}
    r^2 - r - 1 &= 0.
\end{align}

The meaningful solution of this quadratic equation in the interval $1 \leq r \leq 2$ gives the terminal point of the gap:
\begin{align} \label{eq:golden}
    r_{g} &= \frac{1 + \sqrt{5}}{2} \equiv \Phi = 1.618\ldots,
\end{align}
where the golden ratio $\Phi$ is indicated by the vertical dashed orange line in Fig.~\ref{fig:agl_largest_gap}. In this point we obtain $\lambda= \ln{\Phi} = 0.481 \ldots $

 For $r>1$, \eqref{eq:long_term_behavior} is  valid 
 as an unstable extension. 
From it we obtain that
\begin{align}
\lim_{t \to \infty} x_t 
=
\frac{\Phi}{1+\Phi}  
= \frac{1+\sqrt{5}}{3+\sqrt{5}}= 0.618\dots   
\end{align}

From a geometrical perspective it is evident in Fig.~\ref{fig:agl_largest_gap} that 
\begin{align}
    \frac{  \overline{AE} }{ \overline{EB} } =  \frac{  \overline{FG} }{ \overline{GE} }  = \Phi.
\end{align}

Finally, from a dynamical perspective, 
note that there are 3 regimes in Fig.~\ref{fig:agl_largest_gap}, namely:
\begin{itemize}
\item Regular:  for $0 \leq r<1$ the trajectories converge to a fixed point;
\item Chaos with gap: for $1 \leq r< r_{g}$ the trajectories are associated with $\lambda>0$ and are repelled from the regions with gaps;
\item Chaos without gap: for $r \geq r_g$: 
the trajectories are associated with $\lambda>0$ and there is no gap in interval $x_t \in [0,1]$.
\end{itemize}

\begin{figure}[!htb]
\centering
\includegraphics[width=0.498\textwidth]{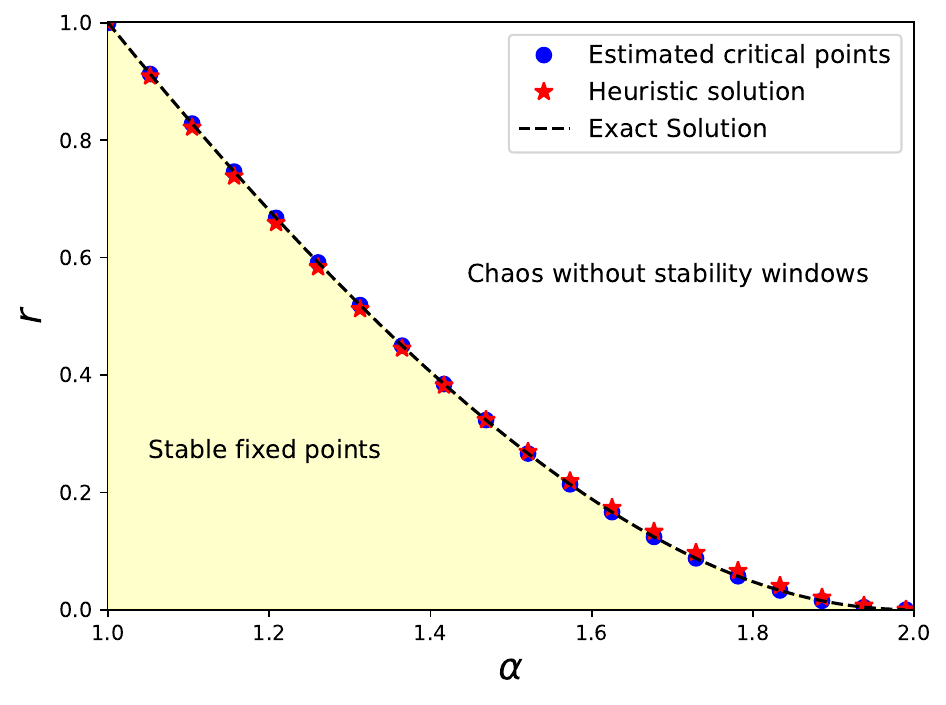}
\caption{Boundary between the regular and chaotic regime for $1 \leq \alpha\leq 2$. The heuristic solution is obtained by a fit with the \eqref{eq:crit_r_heur} where we obtain $ r^*(\alpha) = (2-\alpha)^{1.783} $. The analytical solution comes from \eqref{eq:crit_r_exact}. 
} 

\label{fig:agl_critical_line}
\end{figure}

\begin{figure*}[!htb]
\centering
\includegraphics[scale=0.532]{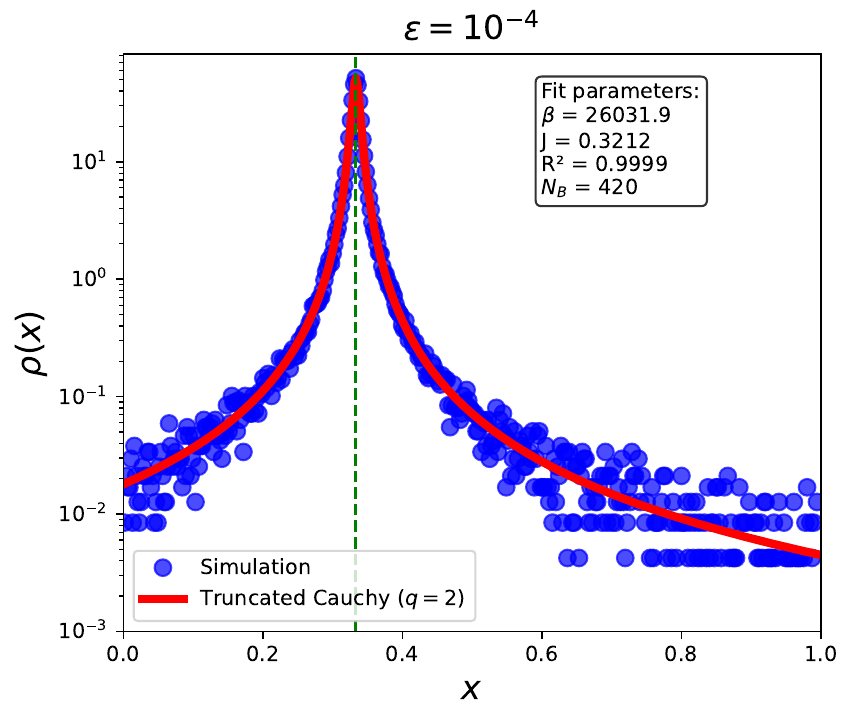}
\includegraphics[scale=0.532]{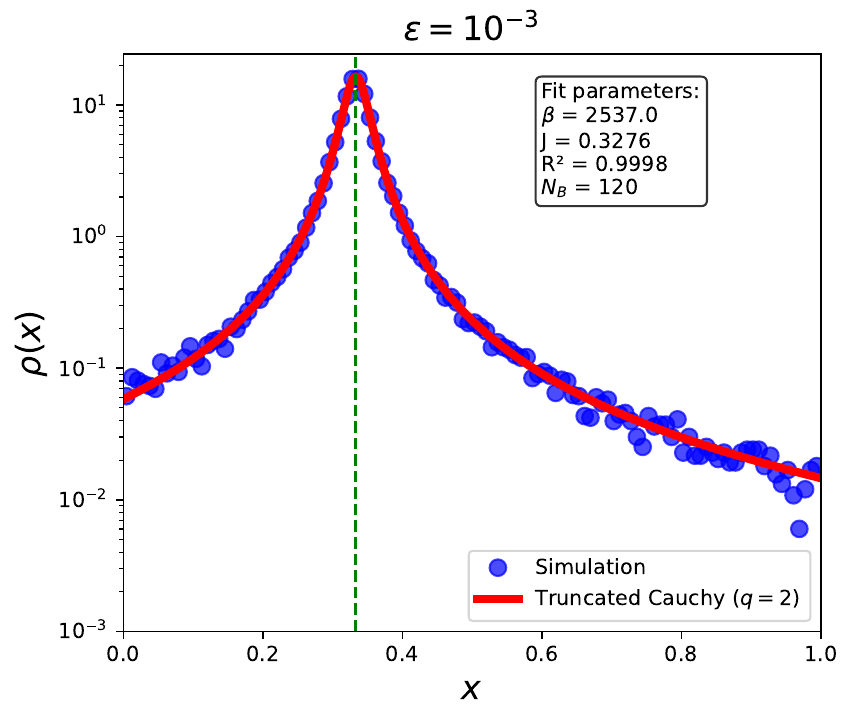}

\includegraphics[scale=0.532]{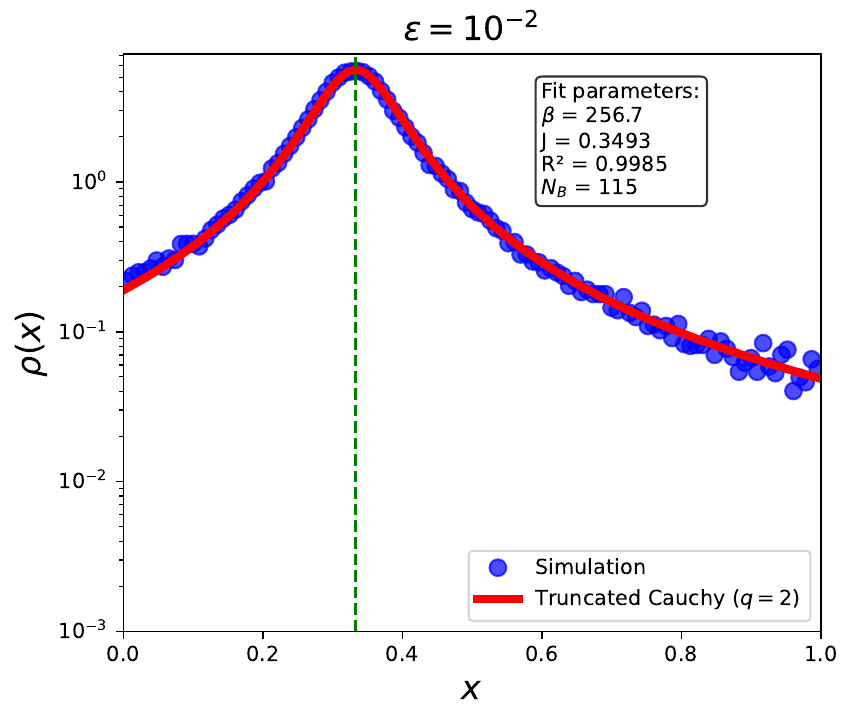}
\includegraphics[scale=0.532]{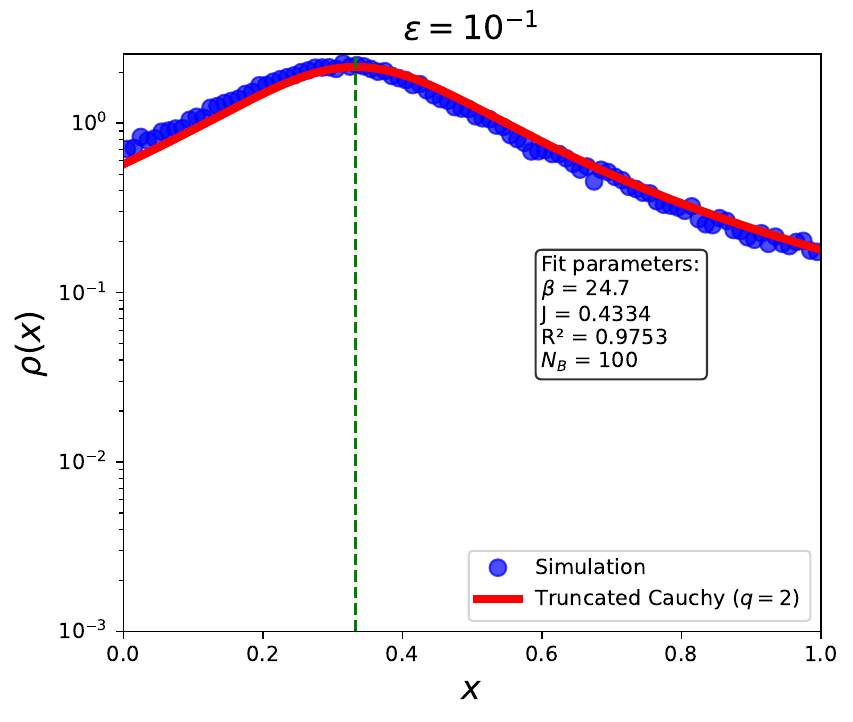}
\caption{
\textcolor{black}{
Invariant density at the edge of the jump to chaos. 
Parameters were set to $\alpha=3/2$ and $r = r^*(1 + \epsilon)$, with $r^*= \sqrt{3}/6 =0.2886...$ established by \eqref{eq:crit_r_exact}. The asymptotic center is located around $x^{*}_c=1/3$, green vertical line, consistent with \eqref{eq:crit_x}. 
Simulation were performed for $t_{\max}=10^5$ timesteps, utilizing $n_{\text{samples}}=10^5$ uniformly distributed initial conditions, $\rho(x_0) \sim U(0,1)$. 
The estimated density, blue points, shows significant agreement with the theoretical truncated Cauchy density, red curve, presented in \eqref{eq:q2_gaussian}.
}
}
\label{fig:edgeofchaos_alp1.5_TCauchy}
\end{figure*}

\subsection{Invariant density}\label{sec:density}

Let $\rho(y)$ denote the invariant density associated with a one-dimensional map $f: [0,1] \to [0,1]$. The Perron-Frobenius equation is given by
\begin{align} \label{eq:pf_general_formal}
    \rho(y) = \sum_{x \in f^{-1}(y)} \frac{\rho(x)}{|f'(x)|}.
\end{align}
From  \eqref{eq:pf_general_formal} we see that we need to analyze the preimages of the map and the corresponding derivatives.

First, in order to gain insights, let us consider $r=3$, see Fig.~\ref{fig:agl_return_and_density_alp1}. The map is defined as $f(x) = 3(1-x) - \lfloor 3(1-x) \rfloor $. This leads to the piecewise linear function:
\begin{align} \label{eq:f_r_3}
f(x) = \begin{cases}
3(1-x) - 0 = 3 - 3x, & 2/3 < x \le 1 \\
3(1-x) - 1 = 2 - 3x, & 1/3 < x \le 2/3 \\
3(1-x) - 2 = 1 - 3x, & 0 \le x \le 1/3.
\end{cases}
\end{align} 
The map consists of $3$ linear pieces, each defined on an interval of length $1/3$. The derivative of $f(x)$ on each piece is $f'(x) = -3$, so $|f'(x)| = 3$ for all $x \in [0,1]$ where the derivative is defined. For any $y \in [0,1)$, there are exactly $3$ preimages $x \in [0,1]$. Let these preimages be $x_1, x_2, x_3$. Applying the Perron-Frobenius equation \eqref{eq:pf_general_formal}, we have
\begin{align} \label{eq:pf_r_3}
    \rho(y) = \frac{\rho(x_1)}{3} + \frac{\rho(x_2)}{3} + \frac{\rho(x_3)}{3}.
\end{align}
This equation is solved for $\rho(y) = 1$, which is exactly the uniform distribution for $y \in [0,1]$.

Now, consider the general case for an integer $r \ge 2$. The map consists of $r$ linear pieces that have equal length $\frac{1}{r}$.
The derivative of $f(x)$ on each interval is $f'(x) = -r$.
The contributions to the invariant density from different parts of the domain are equal, leading to a uniform invariant density $\rho(y) = 1$ for $y \in [0,1)$.

If $r$ is not an integer, the map $f(x) = r(1-x) - \lfloor r(1-x) \rfloor$ continues to have $f'(x) = -r$ in each interval. But the lengths of the domains of these linear segments are no longer the same.  This heterogeneity in the domain lengths leads to a preimage structure that varies with the position within the interval $[0,1]$. Consequently, the invariant density becomes non-uniform.

\color{black}
Figure~\ref{fig:agl_return_and_density_alp1} illustrates the core results of this subsection, with numerical simulations confirming our theoretical predictions. We employ $r=2+\delta$ ($\delta=10^{-6} \ll 1$) rather than $r=2$ because integer values of $r>1$ (particularly even integers) induce numerical instabilities in the dynamics, analogous to what occurs in the tent map $x_{t+1} = A \min(x_t,1-x_t)$ at $A=2$, as documented in Chapter~$2$ of \cite{SprottBook2003}.

In summary,
\begin{align}
    \rho(x) \sim U(0,1) \text{ if $r>1$ is an integer and $\alpha=1$} 
    \label{eq:rho_unif_alp1}
\end{align}
For even $r$, a slight perturbation ($\delta \ll 1$) is essential to avoid numerical artifacts, whereas odd $r$ yields a uniform density, $U(0,1)$,  without such concerns.
\color{black} 

\section{Jump to chaos}\label{sec:jump}

\subsection{Critical line}

From \eqref{eq:explicit_fixed_point} we know that for a given $n^* =  \lfloor g(x^*) \rfloor $ the fixed points of our model satisfy:
\begin{align}
    r (x^*)^{1-\alpha} (1 - x^*) - x^* - n^* &=  0.  \label{eq:explicit_fixed_point2}
\end{align}
The stability of a fixed point is determined by the derivative:
\begin{equation} \label{eq:deriv}
    f'(x^*) = r(1-\alpha)(x^*)^{-\alpha} - r(2-\alpha)(x^*)^{1-\alpha}.
\end{equation}
From a numerical approach, we estimate the critical line using 
\eqref{eq:explicit_fixed_point2} and \eqref{eq:deriv} with the critical point condition $|f'(x^*)| = 1$ for each $\alpha \in [1.0, 1.999]$.

\textcolor{black}{
Figure~\ref{fig:agl_critical_line} shows that   
\eqref{eq:crit_r_exact} accurately captures the critical line. Note that the nature of this transition is not mediated by the Feigenbaum scenario (cascade of period-doubling bifurcations) but instead it manifests as jump to chaos.  
}

\textcolor{black}{As a complementary approach, we now show that a heuristic treatment can also provide insights into the critical line.}
From our main diagram, Fig.~\ref{fig:phase_diagram}, we know that $r^*(\alpha)$  satisfies:
\begin{align}
    r^*(2) &= 0 \label{eq:rcrit_condition_1} \\
    r^*(1) &= 1 \label{eq:rcrit_condition_2}.
\end{align}
From \eqref{eq:rcrit_condition_1} we infer that $r^*(\alpha)= \Psi(2-\alpha)$ where $\Psi$  is a generic function that globally  decays with $\alpha$.
Then, we can assume a two-parameter function  $r^*(\alpha)= a (2-\alpha)^b $. 
Using   \eqref{eq:rcrit_condition_2} we obtain  $r^*(1)= a (1)^b = 1 $, then $a=1$. Thus, consistently with the constraints given by Eqs.~\ref{eq:rcrit_condition_1}-\ref{eq:rcrit_condition_2}, we  
 heuristically approximate the relationship between $\alpha$ and $r^*$ with a single-parameter power-law function:
\begin{equation} \label{eq:crit_r_heur}
    r^*(\alpha) = (2 - \alpha)^b
\end{equation}
Figure~\ref{fig:agl_critical_line} shows that
this ansatz with $b=1.783$ satisfactorily approximates the critical line.

\begin{figure}[!htb]
\centering
\includegraphics[scale=0.70]{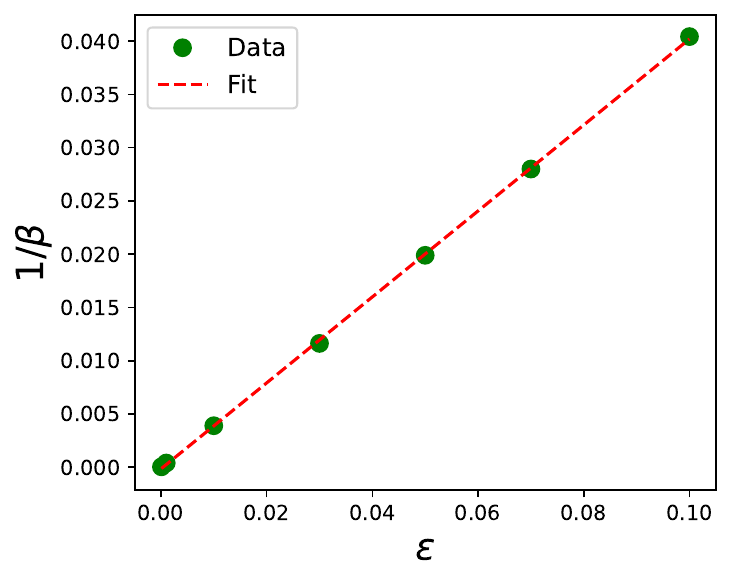}
\caption{ Inverse $\beta$ vs. $\epsilon$. 
 We numerically verify that, for $\alpha \in [1,2]$ and  $0<\epsilon \ll  1 $, $1/\beta =a \, \epsilon$ with $a=0.403$ and $R^2=0.9998$. } 
\label{fig:eps_vs_inverse_beta_TC}
\end{figure}
\begin{figure}[!htb]
    \centering
    \includegraphics[scale=0.70]{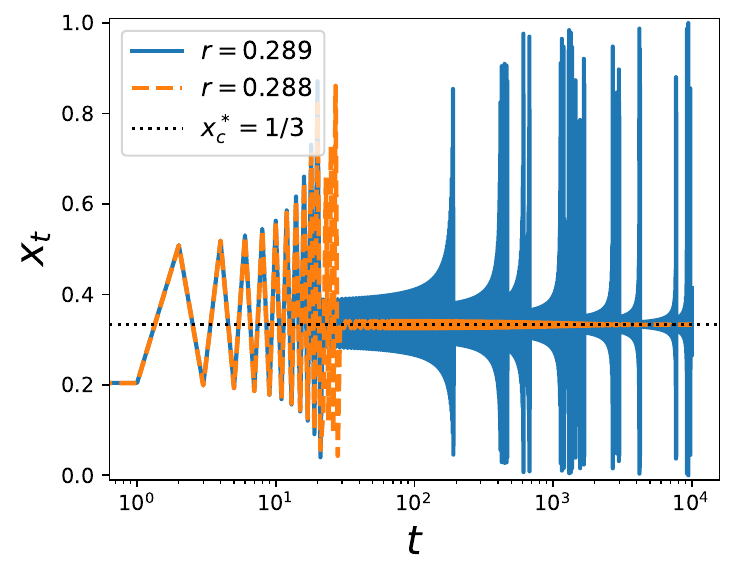} 
\caption{
\textcolor{black}{Time evolution for the $\alpha$GL map with $\alpha=3/2$ for two scenarios: supercritical ($r=0.289 > r^*$) and subcritical ($r=0.288 < r^*$). The critical point, $r^*= \sqrt{3}/6 \approx 0.2886$, is determined by \eqref{eq:crit_r_exact}. The predicted long-time center of the dynamics, obtained from \eqref{eq:crit_x}, is indicated by the dashed black line.
}
}
    \label{fig:edge_timeseries}
\end{figure}

\subsection{Edge of chaos}
 
We investigate the statistical behavior of trajectories generated by the $\alpha$GL map with $\alpha=1.5$. Focusing on the edge of chaos, we analyze the probability distribution by fitting histograms to a truncated $q$-Gaussian~\cite{tsallis2009introduction} with $q = 2$, namely a truncated Cauchy-distribution (also referred  currently to as Cauchy-Lorentz distribution)  
\begin{align} \label{eq:q2_gaussian}
\rho(x)=J\frac{ \sqrt{\beta} }{1+\beta (x-x^*_c)^2}, 
\end{align}
where $x^*_c$ comes from \eqref{eq:crit_x}  and $J$ is a normalization constant that is computed numerically, subject to the constraint $x \in [0,1]$. For unconstrained dynamics (i.e., without this range restriction), the normalization simplifies to $J=1/\pi$.

The quality of the fit is evaluated through the coefficient of determination ($R^2$). The optimal parameter set $\{\beta, J\}$ is identified by selecting the best fit across varying numbers of bins ($N_B$) and choosing the configuration that yields the maximal $R^2$ value.

Figure~\ref{fig:edgeofchaos_alp1.5_TCauchy} shows that the $q$-Gaussian with $q=2$ effectively captures the statistical behavior at the edge of chaos, providing further support to the conjecture of universality of the $q=2$-Gaussian shape of the invariant density close to the critical point in a jump to chaos~\cite{Beck2024}.

Figure~\ref{fig:eps_vs_inverse_beta_TC} strongly indicates that 
$ \lim_{\epsilon \to 0}1/\beta =0 $ . This is expected since the limit behavior should be marked by a delta-like invariant density associated with the exact boundary between the fixed-point regime and chaos.

Figure~\ref{fig:edge_timeseries} indicates that the temporal evolution at the edge of the jump to chaos exhibits two competing effects: sudden returns toward the center and a gradual exponential-like separation from the center of the distribution. These competing mechanisms repeat intermittently, producing the observed long tail in the invariant density close to the critical point of the abrupt transition to chaos.\\

\section{Final remarks}\label{sec:final}

\textcolor{black}{
This work was motivated by a fundamental inquiry into the gradual and sudden routes to chaos, which is an important topic in the foundations of chaos theory. Also, this map has potential applications to biological-like systems due to the nature of the logistic map.
}
While the logistic map and the recently introduced $\alpha$-Gauss map have been studied separately, our work couples these models for the first time in what we call the $\alpha$-Gauss-Logistic  ($\alpha$GL) map. Our model exhibits a rich phenomenology.

\textcolor{black}{
For $\alpha<1$ the $\alpha$GL map undergoes multiple sequential bifurcation cascades. For each cascade we note a Feigenbaum-like scenario with progressive period-doubling until reaching a chaotic regime.
For $1 \leq \alpha <2 $, the system displays an abrupt transition
from the regular regime to chaos without  period-doubling stages. 
}

We find a manifestation of robust chaos~\cite{Banerjee1998,ZeraouliaSprott2012} for  $\alpha \geq 1$, characterized by an absence of islands of stability. This feature is not present for $\alpha<1$ where the chaotic attractor is interspersed by periodic windows and exhibits a nonmonotonic Lyapunov exponent.

\textcolor{black}{
In the case where $\alpha = 0$, the proposed model transforms into an \emph{extended logistic map}.
This new model features a parameter $r$ that is no longer confined to a specific interval, yet $x_t$ remain within the unit interval. This distinguishes it from prior generalizations of the logistic map \cite{radwan2013some,borujeni2015modified,lawnik2017generalized,da2017route,sayed2017generalized,leonel2019allee,hamada2025investigating,zhang2025chaos,abdellah2025generalized}. 
For our extended logistic map, we have successfully obtained analytical solutions that precisely delineate the parameter region of $r$ where both fixed points and 2-cycles can be located.
}

For the special case $\alpha = 1$, we obtain several analytical results including an explicit expression for the Lyapunov exponent given by 
$\lambda_{\alpha=1} = \ln r$. A complete regime diagram for $\alpha = 1$ shows  chaotic phases with or without gaps. We prove the unexpected  emergence of the golden ratio $\Phi$ as the parameter threshold marking the end of the largest prominent gap in the $\alpha = 1$ regime diagram.  

Using the Perron-Frobenius equation we demonstrate that for an odd integer $r>1$ and $\alpha=1$ our map exhibits an exact uniform density. This is different from Ref.~\cite{Beck2024} where the uniform  density  is approached for $\alpha \to \infty$ in the 
$\alpha$-Gauss map.

At the edge of chaos, in the sudden transition region, we identify a Cauchy distribution ($q=2$) for the invariant measure, supporting a recent conjecture~\cite{Beck2024} that systems exhibiting abrupt transitions to chaos display $q$-Gaussian distributions at the boundary between fixed-point and chaotic regimes. \textcolor{black}{From a broader point of view our results provide a novel model in which $q$-Gaussians are observed. For instance, another map that exhibits this type of distribution at the edge of chaos is the 
two–dimensional standard map as it was numerically verified in 
Ref.~\cite{tirnakli2016standard} and analytically proved in Ref.~\cite{bountis2020cauchy}. }

In future works, it might be interesting to characterize the patterns of the time series produced by various routes to chaos using novel methodologies like the binary complexity-entropy plane~\cite{pinto2025cryptocurrency}. Additionally,  \textcolor{black}{it would be   interesting to analyze high-dimensional versions of our $\alpha$GL map.}

\section*{Acknowledgements}
Two of us (CT and EMFC) acknowledge partial financial support from CNPq and Faperj (Brazilian Agencies).

\newpage
\bibliography{main.bib}

\end{document}